\documentclass{article}%
\usepackage{amsfonts}
\usepackage{amssymb}
\usepackage{amsmath}
\usepackage{graphicx}%
\newenvironment{proof}[1][Proof]{\noindent\textbf{#1.} }{\ \rule{0.5em}{0.5em}}
\begin{document}

\author{{\small V. V. Fern\'{a}ndez}$^{{\footnotesize 1}}${\small , A. M.
Moya}$^{{\footnotesize 1}}${\small , E. Notte-Cuello}$^{{\footnotesize 2}}%
${\small \ and W. A. Rodrigues Jr.}$^{{\footnotesize 1}}${\small . }\\$^{{\footnotesize 1}}\hspace{-0.1cm}${\footnotesize Institute of Mathematics,
Statistics and Scientific Computation}\\{\footnotesize IMECC-UNICAMP CP 6065}\\{\footnotesize 13083-859 Campinas, SP, Brazil}\\$^{{\footnotesize 2}}${\small Departamento de} {\small Matem\'{a}ticas,}\\{\small Universidad de La Serena}\\{\small Av. Cisternas 1200, La Serena-Chile}\\{\small e-mail:} {\small walrod@ime.unicamp.br and enotte@userena.cl }}
\title{Covariant Derivatives of Multivector and Multiform Fields}
\maketitle

\begin{abstract}
A simple theory of the covariant derivatives, deformed derivatives and
relative covariant derivatives of multivector and multiform fields is
presented using algebraic and analytical tools developed in previous papers.

\end{abstract}
\tableofcontents

\section{Introduction}

Using an arbitrary parallelism structure $\langle U,\Gamma\rangle$ on an open
set $U\ $\ of a smooth manifold $M$ as defined in \cite{fmcr2} we present,
using the algebraic tools developed in \cite{qr07,fmcr1} a detailed theory of
the covariant derivatives, deformed covariant derivatives and relative
covariant derivatives of \textit{multivector} and \textit{multiform} fields
(objects that represent important fields used in physical theories). Several
useful formulas needed for practical calculations are derived.

\section{Multivector and Multiform Fields}

Let $U$ be an open set \ on the smooth finite dimensional manifold $M$ (i.e.,
$\dim M=n$ with $n\in\mathbb{N}$). As we know, the set of smooth\footnote{In
this paper smooth means $\mathcal{C}^{\infty}$-differentiable or at least
enough differentiable for our statements to hold.} scalar fields on $U$ has
natural structure of \emph{ring }(\emph{with unity}). It will be denoted by
$\mathcal{S}(U)$. The set of smooth vector fields on $U$ has natural structure
of \emph{module over the ring} $\mathcal{S}(U)$. It will be denoted by
$\mathcal{V}(U)$. The set of smooth form fields on $U$ \emph{can be
identified} with the \emph{dual module} for $\mathcal{V}(U)$. It could be
denoted by $\mathcal{V}^{\ast}(U)$.

A $k$-vector mapping%
\begin{equation}
X^{k}:U\longrightarrow\underset{p\in U}{%
%TCIMACRO{\tbigcup }%
%BeginExpansion
{\textstyle\bigcup}
%EndExpansion
}%
%TCIMACRO{\tbigwedge \nolimits^{k}}%
%BeginExpansion
{\textstyle\bigwedge\nolimits^{k}}
%EndExpansion
T_{p}^{\ast}M,
\end{equation}
such that for each $p\in U$, $X_{(p)}^{k}\in%
%TCIMACRO{\tbigwedge \nolimits^{k}}%
%BeginExpansion
{\textstyle\bigwedge\nolimits^{k}}
%EndExpansion
T_{p}^{\ast}M$ is called a $k$\emph{-vector field on }$U.$

Such $X^{k}$ with $1\leq k\leq n$ is said to be a smooth $k$-vector field on
$U,$ if and only if, for all $\omega^{1},\ldots,\omega^{k}\in\mathcal{V}%
^{\ast}(U)$, the scalar mapping defined by%
\begin{equation}
U\ni p\longmapsto X_{(p)}^{k}(\omega_{(p)}^{1},\ldots,\omega_{(p)}^{k}%
)\in\mathbb{R} \label{MMF2}%
\end{equation}
is a smooth scalar field on $U.$

A multivector mapping%
\begin{equation}
X:U\longrightarrow\underset{p\in U}{%
%TCIMACRO{\tbigcup }%
%BeginExpansion
{\textstyle\bigcup}
%EndExpansion
}%
%TCIMACRO{\tbigwedge }%
%BeginExpansion
{\textstyle\bigwedge}
%EndExpansion
T_{p}^{\ast}M,
\end{equation}
such that for each $p\in U$, $X_{(p)}\in%
%TCIMACRO{\tbigwedge }%
%BeginExpansion
{\textstyle\bigwedge}
%EndExpansion
T_{p}^{\ast}M$ is called a \emph{multivector field on }$U.$

Any multivector at $p\in M$ can be written as a sum of $k$-vectors (i.e.,
homogeneous multivectors of degree $k$) at $p\in M,$ with $k$ running from
$k=0$ to $k=n$, i.e., there exist exactly $n+1$ homogeneous multivector of
degree $k$ fields on $U,$ conveniently denote by $X^{0},X^{1},\ldots,X^{n},$
such that for every $p\in U,$%
\begin{equation}
X_{(p)}=X_{(p)}^{0}+X_{(p)}^{1}+\cdots+X_{(p)}^{n}. \label{MMF4}%
\end{equation}

We say that $X$ is a smooth multivector field on $U$ when each of one of the
$X^{0},X^{1},\ldots,X^{n}$ is a smooth $k$-vector field on $U.$

We emphasize that according with the definitions of smoothness as given above,
a smooth $k$-vector field on $U$ \emph{can be identified} to a $k$-vector over
$\mathcal{V}(U)$, and a smooth multivector field on $U$ \emph{can be seen
properly} as a multivector over $\mathcal{V}(U).$ Thus, the set of smooth
$k$-vector fields on $U$ may be denoted by $%
%TCIMACRO{\tbigwedge \nolimits^{k}}%
%BeginExpansion
{\textstyle\bigwedge\nolimits^{k}}
%EndExpansion
\mathcal{V}(U)$, and the set of smooth multivector fields on $U$ may be
denoted by $%
%TCIMACRO{\tbigwedge }%
%BeginExpansion
{\textstyle\bigwedge}
%EndExpansion
\mathcal{V}(U)$.

A $k$-form mapping%
\begin{equation}
\Phi_{k}:U\longrightarrow\underset{p\in U}{%
%TCIMACRO{\tbigcup }%
%BeginExpansion
{\textstyle\bigcup}
%EndExpansion
}%
%TCIMACRO{\tbigwedge \nolimits^{k}}%
%BeginExpansion
{\textstyle\bigwedge\nolimits^{k}}
%EndExpansion
T_{p}^{\ast}M
\end{equation}
such that for each $p\in U$, $\Phi_{k(p)}\in%
%TCIMACRO{\tbigwedge \nolimits^{k}}%
%BeginExpansion
{\textstyle\bigwedge\nolimits^{k}}
%EndExpansion
T_{p}^{\ast}M$ is called a $k$\emph{-form field on }$U.$

Such $\Phi_{k}$ with $1\leq k\leq n$ is said to be a smooth $k$-form field on
$U,$ if and only if, for all $v_{1},\ldots,v_{k}\in\mathcal{V}(U)$, the scalar
mapping defined by%
\begin{equation}
U\ni p\longmapsto\Phi_{k(p)}(v_{1(p)},\ldots,v_{k(p)})\in\mathbb{R}
\label{MMF6}%
\end{equation}
is a smooth scalar field on $U$.

A multiform mapping%
\begin{equation}
\Phi:U\longrightarrow\underset{p\in U}{%
%TCIMACRO{\tbigcup }%
%BeginExpansion
{\textstyle\bigcup}
%EndExpansion
}%
%TCIMACRO{\tbigwedge }%
%BeginExpansion
{\textstyle\bigwedge}
%EndExpansion
T_{p}^{\ast}M
\end{equation}
such that for each $p\in U$, $\Phi_{(p)}\in%
%TCIMACRO{\tbigwedge }%
%BeginExpansion
{\textstyle\bigwedge}
%EndExpansion
T_{p}^{\ast}M$ is called a \emph{multiform field on }$U.$

Any multiform at $p\in M$ can be written (see, e.g.,\cite{rodoliv2006}) as a
sum of $k$-forms (i.e., homogeneous multiforms of degree $k$) at $p\in M$ with
$k$ running from $k=0$ to $k=n.$ It follows that there exist exactly $n+1$
homogeneous multiform of degree $k$ fields on $U,$ named as $\Phi_{0},\Phi
_{1},\ldots,\Phi_{n}$ such that%
\begin{equation}
\Phi_{(p)}=\Phi_{0(p)}+\Phi_{1(p)}+\cdots+\Phi_{n(p)} \label{MMF8}%
\end{equation}
for every $p\in U.$

We say that $\Phi$ is a smooth multiform field on $U,$ if and only if, each of
$\Phi_{0},\Phi_{1},\ldots,\Phi_{n}$ is just a smooth $k$-form field on $U.$

Note that according with the definitions of smoothness as given above, a
smooth $k$-form field on $U$ \emph{can be identified} with a $k$-form over
$\mathcal{V}(U),$ and a smooth multiform field on $U$ \emph{can be seen} as a
multiform over $\mathcal{V}(U).$ Thus, the set of smooth $k$-form fields on
$U$ will be denoted by $%
%TCIMACRO{\tbigwedge \nolimits^{k}}%
%BeginExpansion
{\textstyle\bigwedge\nolimits^{k}}
%EndExpansion
\mathcal{V}^{\ast}(U),$ and the set of smooth multiform fields on $U$ will be
denoted by $%
%TCIMACRO{\tbigwedge }%
%BeginExpansion
{\textstyle\bigwedge}
%EndExpansion
\mathcal{V}^{\ast}(U).$

\subsection{Algebras of Multivector and Multiform Fields}

We recall first the module (over a ring) structure operations of the set of
smooth multivector fields on $U$ and of the set of smooth multiform fields on
$U$. We recall also the concept of the exterior product both of smooth
multivector fields as well as of smooth multiform fields on $U$. Finally, we
present the definitions of the duality products of a given smooth multivector
fields on $U$ by a smooth multiform field on $U.$

The addition of multivector fields $X$ and $Y,$ or multiform fields $\Phi$ and
$\Psi,$ is defined by%
\begin{align}
\left(  X+Y\right)  _{(p)}  &  =X_{(p)}+Y_{(p)},\label{MMF9}\\
\left(  \Phi+\Psi\right)  _{(p)}  &  =\Phi_{(p)}+\Psi_{(p)}, \label{MMF10}%
\end{align}
for every $p\in U.$

The scalar multiplication of a multivector field $X,$ or a multiform field
$\Phi,$ by a scalar field $f,$ is defined by%
\begin{align}
\left(  fX\right)  _{(p)}  &  =f(p)X_{(p)},\label{MMF11}\\
\left(  f\Phi\right)  _{(p)}  &  =f(p)\Phi_{(p)}, \label{MMF12}%
\end{align}
for every $p\in U.$

The exterior product of multivector fields $X$ and $Y,$ and the exterior
product of multiform fields $\Phi$ and $\Psi,$ are defined by%
\begin{align}
\left(  X\wedge Y\right)  _{(p)}  &  =X_{(p)}\wedge Y_{(p)},\label{MMF13}\\
\left(  \Phi\wedge\Psi\right)  _{(p)}  &  =\Phi_{(p)}\wedge\Psi_{(p)},
\label{MMF14}%
\end{align}
for every $p\in U.$

Each module, of either the smooth multivector fields on $U$, or the smooth
multiform fields on $U$, endowed with the respective exterior product has a
natural structure of associative algebra. They are called \emph{the exterior
algebras of multivector and multiform fields on} $U.$

The duality scalar product of a multiform field $\Phi$ with a multivector
field $X$ is (see algebraic details in \cite{fmcr1}) the scalar field
$\left\langle \Phi,X\right\rangle $ defined by%
\begin{equation}
\left\langle \Phi,X\right\rangle (p)=\left\langle \Phi_{(p)},X_{(p)}%
\right\rangle , \label{MMF15}%
\end{equation}
for every $p\in U.$

The duality left contracted product of a multiform field $\Phi$ with a
multivector field $X$ (or, a multivector field $X$ with a multiform field
$\Phi$) is the multivector field $\left\langle \Phi,X\right\vert $
(respectively, the multiform field $\left\langle X,\Phi\right\vert $) defined
by%
\begin{align}
\left\langle \Phi,X\right\vert _{(p)}  &  =\left\langle \Phi_{(p)}%
,X_{(p)}\right\vert ,\label{MMF16}\\
\left\langle X,\Phi\right\vert _{(p)}  &  =\left\langle X_{(p)},\Phi
_{(p)}\right\vert , \label{MMF17}%
\end{align}
for every $p\in U.$

The duality right contracted product of a multiform field $\Phi$ with a
multivector field $X$ (or, a multivector field $X$ with a multiform field
$\Phi$) is the multiform field $\left\vert \Phi,X\right\rangle $
(respectively, the multivector field $\left\vert X,\Phi\right\rangle $)
defined by%
\begin{align}
\left\vert \Phi,X\right\rangle _{(p)}  &  =\left\vert \Phi_{(p)}%
,X_{(p)}\right\rangle ,\label{MMF18}\\
\left\vert X,\Phi\right\rangle _{(p)}  &  =\left\vert X_{(p)},\Phi
_{(p)}\right\rangle , \label{MMF19}%
\end{align}
for every $p\in U.$

Each duality contracted product of smooth multivector fields on $U$ with
smooth multiform fields on $U$ yields a natural structure of
\textit{non-associative} algebra.

\section{Covariant Derivative of Multivector and Multiform Fields}

Let $\left\langle U,\Gamma\right\rangle $ be a parallelism structure
\cite{fmcr2} on $U,$ and let $a\in\mathcal{V}(U)$. The $\emph{a}%
$\emph{-Directional Covariant Derivative} ($\emph{a}$\emph{-DCD})\emph{\ }of a
smooth multivector field on $U,$ associated with $\left\langle U,\Gamma
\right\rangle $, is the mapping%
\[%
%TCIMACRO{\dbigwedge }%
%BeginExpansion
{\displaystyle\bigwedge}
%EndExpansion
\mathcal{V}(U)\ni X\longmapsto\nabla_{a}X\in%
%TCIMACRO{\dbigwedge }%
%BeginExpansion
{\displaystyle\bigwedge}
%EndExpansion
\mathcal{V}(U),
\]
such that the following axioms are satisfied:

\textbf{i. }For all $f\in\mathcal{S}(U):$%
\begin{equation}
\nabla_{a}f=af. \label{CDMMF1}%
\end{equation}

\textbf{ii.} For all $X^{k}\in%
%TCIMACRO{\tbigwedge \nolimits^{k}}%
%BeginExpansion
{\textstyle\bigwedge\nolimits^{k}}
%EndExpansion
\mathcal{V}(U)$ with $k\geq1:$%
\begin{align}
\nabla_{a}X^{k}(\omega^{1},\ldots,\omega^{k})  &  =aX^{k}(\omega^{1}%
,\ldots,\omega^{k})\nonumber\\
&  -X^{k}(\nabla_{a}\omega^{1},\ldots,\omega^{k})\cdots-X^{k}(\omega
^{1},\ldots,\nabla_{a}\omega^{k}), \label{CDMMF2}%
\end{align}
for every $\omega^{1},\ldots,\omega^{k}\in\mathcal{V}^{\star}(U).$

\textbf{iii.} For all $X\in%
%TCIMACRO{\tbigwedge }%
%BeginExpansion
{\textstyle\bigwedge}
%EndExpansion
\mathcal{V}(U)$, if $X=\overset{n}{\underset{k=0}{%
%TCIMACRO{\tsum }%
%BeginExpansion
{\textstyle\sum}
%EndExpansion
}}X^{k}$, then%
\begin{equation}
\nabla_{a}X=\overset{n}{\underset{k=0}{%
%TCIMACRO{\tsum }%
%BeginExpansion
{\textstyle\sum}
%EndExpansion
}}\nabla_{a}X^{k}. \label{CDMMF3}%
\end{equation}

The basic properties of the $a$-\textit{DCD} of smooth multivector fields are:

\begin{itemize}
\item The $a$-Directional Covariant Derivative Operator $\nabla_{a}$ when
acting on multivector fields is grade-preserving, i.e.,
\begin{equation}
\text{if }X\in%
%TCIMACRO{\tbigwedge \nolimits^{k}}%
%BeginExpansion
{\textstyle\bigwedge\nolimits^{k}}
%EndExpansion
\mathcal{V}(U),\text{ then }\nabla_{a}X\in%
%TCIMACRO{\tbigwedge \nolimits^{k}}%
%BeginExpansion
{\textstyle\bigwedge\nolimits^{k}}
%EndExpansion
\mathcal{V}(U). \label{CDMMF4}%
\end{equation}

\item \textbf{\ }For all $f\in\mathcal{S}(U),$ $a,b\in\mathcal{V}(U)$ and
$X\in%
%TCIMACRO{\dbigwedge }%
%BeginExpansion
{\displaystyle\bigwedge}
%EndExpansion
\mathcal{V}(U)$%
\begin{align}
\nabla_{a+b}X  &  =\nabla_{a}X+\nabla_{b}X,\nonumber\\
\nabla_{fa}X  &  =f\nabla_{a}X. \label{CDMMF5}%
\end{align}

\item \textbf{\ }For all $f\in\mathcal{S}(U),$ $a\in\mathcal{V}(U)$ and
$X,Y\in%
%TCIMACRO{\dbigwedge }%
%BeginExpansion
{\displaystyle\bigwedge}
%EndExpansion
\mathcal{V}(U)$%
\begin{align}
\nabla_{a}(X+Y)  &  =\nabla_{a}X+\nabla_{a}Y,\nonumber\\
\nabla_{a}(fX)  &  =(af)X+f\nabla_{a}X. \label{CDMMF6}%
\end{align}

\item For all $a\in\mathcal{V}(U)$ and $X,Y\in%
%TCIMACRO{\dbigwedge }%
%BeginExpansion
{\displaystyle\bigwedge}
%EndExpansion
\mathcal{V}(U)$%
\begin{equation}
\nabla_{a}(X\wedge Y)=(\nabla_{a}X)\wedge Y+X\wedge\nabla_{a}Y. \label{CDMMF7}%
\end{equation}

\end{itemize}

The $\emph{a}$\emph{-}Directional Covariant Derivative of a smooth multiform
field on $U$ associated with $\left\langle U,\Gamma\right\rangle $ is the
mapping%
\[%
%TCIMACRO{\dbigwedge }%
%BeginExpansion
{\displaystyle\bigwedge}
%EndExpansion
\mathcal{V}^{\star}(U)\ni\Phi\longmapsto\nabla_{a}\Phi\in%
%TCIMACRO{\dbigwedge }%
%BeginExpansion
{\displaystyle\bigwedge}
%EndExpansion
\mathcal{V}^{\star}(U),
\]
such that the following axioms are satisfied:

\textbf{\ i. }For all $f\in\mathcal{S}(U):$%
\begin{equation}
\nabla_{a}f=af. \label{CDMMF8}%
\end{equation}

\textbf{ii.} For all $\Phi_{k}\in%
%TCIMACRO{\dbigwedge ^{k}}%
%BeginExpansion
{\displaystyle\bigwedge^{k}}
%EndExpansion
\mathcal{V}^{\star}(U)$ with $k\geq1:$%
\begin{align}
\nabla_{a}\Phi_{k}(v_{1},\ldots,v_{k})  &  =a\Phi_{k}(v_{1},\ldots
,v_{k})\nonumber\\
&  -\Phi_{k}(\nabla_{a}v_{1},\ldots,v_{k})\cdots-\Phi_{k}(v_{1},\ldots
,\nabla_{a}v_{k}), \label{CDMMF9}%
\end{align}
for every $v_{1},\ldots,v_{k}\in\mathcal{V}(U).$

\textbf{iii.} For all $\Phi\in%
%TCIMACRO{\dbigwedge }%
%BeginExpansion
{\displaystyle\bigwedge}
%EndExpansion
\mathcal{V}^{\star}(U):$ if $\Phi=\underset{k=0}{\overset{n}{%
%TCIMACRO{\tsum }%
%BeginExpansion
{\textstyle\sum}
%EndExpansion
}}\Phi_{k},$ then%
\begin{equation}
\nabla_{a}\Phi=\underset{k=0}{\overset{n}{%
%TCIMACRO{\tsum }%
%BeginExpansion
{\textstyle\sum}
%EndExpansion
}}\nabla_{a}\Phi_{k}. \label{CDMMF10}%
\end{equation}

The basic properties for the $a$-\textit{DCD} of smooth multiform fields.

\begin{itemize}
\item \textbf{\ }The $a$-\textit{DCD} $\nabla_{a}$ when acting on multiform
fields is grade-preserving, i.e.,
\begin{equation}
\text{if }\Phi\in%
%TCIMACRO{\tbigwedge \nolimits^{k}}%
%BeginExpansion
{\textstyle\bigwedge\nolimits^{k}}
%EndExpansion
\mathcal{V}^{\star}(U),\text{ then }\nabla_{a}\Phi\in%
%TCIMACRO{\tbigwedge \nolimits^{k}}%
%BeginExpansion
{\textstyle\bigwedge\nolimits^{k}}
%EndExpansion
\mathcal{V}^{\star}(U). \label{CDMMF11}%
\end{equation}

\item \textbf{\ }For all $f\in\mathcal{S}(U),$ $a,b\in\mathcal{V}(U)$ and
$\Phi\in%
%TCIMACRO{\dbigwedge }%
%BeginExpansion
{\displaystyle\bigwedge}
%EndExpansion
\mathcal{V}^{\star}(U)$
\begin{align}
\nabla_{a+b}\Phi &  =\nabla_{a}\Phi+\nabla_{b}\Phi,\nonumber\\
\nabla_{fa}\Phi &  =f\nabla_{a}\Phi. \label{CDMMF12}%
\end{align}

\item \textbf{\ }For all $f\in\mathcal{S}(u),$ $a\in\mathcal{V}(U)$ and
$\Phi,\Psi\in%
%TCIMACRO{\dbigwedge }%
%BeginExpansion
{\displaystyle\bigwedge}
%EndExpansion
\mathcal{V}^{\star}(U)$
\begin{align}
\nabla_{a}(\Phi+\Psi)  &  =\nabla_{a}\Phi+\nabla_{a}\Psi,\nonumber\\
\nabla_{a}(f\Phi)  &  =(af)\Phi+f\nabla_{a}\Phi. \label{CDMMF13}%
\end{align}

\item \textbf{\ }For all $a\in\mathcal{V}(U)$ and $\Phi,\Psi\in%
%TCIMACRO{\dbigwedge }%
%BeginExpansion
{\displaystyle\bigwedge}
%EndExpansion
\mathcal{V}^{\star}(U)$
\begin{equation}
\nabla_{a}(\Phi\wedge\Psi)=(\nabla_{a}\Phi)\wedge\Psi+\Phi\wedge\nabla_{a}%
\Psi. \label{CDMMF14}%
\end{equation}
\medskip
\end{itemize}

We now present three remarkable properties involving the action of $\nabla
_{a}$ on the duality products of multivector and multiform fields.

\begin{itemize}
\item When $\nabla_{a}$ acts on the duality scalar product of $\Phi\in%
%TCIMACRO{\dbigwedge }%
%BeginExpansion
{\displaystyle\bigwedge}
%EndExpansion
\mathcal{V}^{\star}(U)$ with $X\in%
%TCIMACRO{\dbigwedge }%
%BeginExpansion
{\displaystyle\bigwedge}
%EndExpansion
\mathcal{V}(U)$ follows the Leibniz rule, i.e.,%
\begin{equation}
a\left\langle \Phi,X\right\rangle =\left\langle \nabla_{a}\Phi,X\right\rangle
+\left\langle \Phi,\nabla_{a}X\right\rangle .\label{CDMMF15}%
\end{equation}

\end{itemize}

\begin{proof}
We check this statement only for the duality scalar product of $\omega
\in\mathcal{V}^{\star}(U)$ with $v\in\mathcal{V}(U),$ and of $\Phi_{2}\in%
%TCIMACRO{\dbigwedge ^{2}}%
%BeginExpansion
{\displaystyle\bigwedge^{2}}
%EndExpansion
\mathcal{V}^{\star}(U)$ with $X^{2}\in%
%TCIMACRO{\dbigwedge ^{2}}%
%BeginExpansion
{\displaystyle\bigwedge^{2}}
%EndExpansion
\mathcal{V}(U)$.

For $\omega\in\mathcal{V}^{\star}(U)$ and $v\in\mathcal{V}(U),$ using
Eq.(\ref{CDMMF9}), we have%
\begin{equation}
\nabla_{a}\omega(v)=a\omega(v)-\omega(\nabla_{a}v), \tag{a}%
\end{equation}
but, according to with Eq.(\ref{MMF15}), it can be written%
\begin{equation}
a\left\langle \omega,v\right\rangle =\left\langle \nabla_{a}\omega
,v\right\rangle +\left\langle \omega,\nabla_{a}v\right\rangle . \tag{b}%
\end{equation}

For $\Phi_{2}\in%
%TCIMACRO{\dbigwedge ^{2}}%
%BeginExpansion
{\displaystyle\bigwedge^{2}}
%EndExpansion
\mathcal{V}^{\star}(U)$ and $X^{2}\in%
%TCIMACRO{\dbigwedge ^{2}}%
%BeginExpansion
{\displaystyle\bigwedge^{2}}
%EndExpansion
\mathcal{V}(U),$ the definition of duality scalar product holds%
\begin{equation}
\left\langle \Phi_{2},X^{2}\right\rangle =\frac{1}{2!}\Phi_{2}(e_{j}%
,e_{k})X^{2}(\varepsilon^{j},\varepsilon^{k}), \tag{c}%
\end{equation}
where $\left\{  e_{j},\varepsilon^{j}\right\}  $ is any \emph{pair of dual
frame fields on }$U$.

A straightforward calculation, using Eq.(\ref{CDMMF9}), yields%
\begin{align}
a\left\langle \Phi_{2},X^{2}\right\rangle  &  =\frac{1}{2!}(a\Phi_{2}%
(e_{j},e_{k})X^{2}(\varepsilon^{j},\varepsilon^{k})+\Phi_{2}(e_{j}%
,e_{k})aX^{2}(\varepsilon^{j},\varepsilon^{k}))\nonumber\\
&  =\frac{1}{2!}\nabla_{a}\Phi_{2}(e_{j},e_{k})X^{2}(\varepsilon
^{j},\varepsilon^{k})+\frac{1}{2}\Phi_{2}(\nabla_{a}e_{j},e_{k})X^{2}%
(\varepsilon^{j},\varepsilon^{k})\nonumber\\
&  +\frac{1}{2}\Phi_{2}(e_{j},\nabla_{a}e_{k})X^{2}(\varepsilon^{j}%
,\varepsilon^{k})+\frac{1}{2}\Phi_{2}(e_{j},e_{k})X^{2}(\varepsilon^{j}%
,\nabla_{a}\varepsilon^{k})\tag{d}\\
&  +\frac{1}{2}\Phi_{2}(e_{j},e_{k})X^{2}(\nabla_{a}\varepsilon^{j}%
,\varepsilon^{k})+\frac{1}{2!}\Phi_{2}(e_{j},e_{k})\nabla_{a}X^{2}%
(\varepsilon^{j},\varepsilon^{k}),\nonumber
\end{align}
and recalling (b), we get%
\begin{align}
\Phi_{2}(\nabla_{a}e_{j},e_{k})X^{2}(\varepsilon^{j},\varepsilon^{k})  &
=\Phi_{2}(\left\langle \varepsilon^{p},\nabla_{a}e_{j}\right\rangle
e_{p},e_{k})X^{2}(\varepsilon^{j},\varepsilon^{k})\nonumber\\
&  =\Phi_{2}(e_{p},e_{k})X^{2}(\left\langle \varepsilon^{p},\nabla_{a}%
e_{j}\right\rangle \varepsilon^{j},\varepsilon^{k})\nonumber\\
&  =\Phi_{2}(e_{p},e_{k})X^{2}(-\left\langle \nabla_{a}\varepsilon^{p}%
,e_{j}\right\rangle \varepsilon^{j},\varepsilon^{k})\nonumber\\
&  =-\Phi_{2}(e_{p},e_{k})X^{2}(\nabla_{a}\varepsilon^{p},\varepsilon^{k}).
\tag{e}%
\end{align}

Thus, by putting (e) into (d), and by using once again the definition of
duality scalar product, we finally have%
\begin{equation}
a\left\langle \Phi_{2},X^{2}\right\rangle =\left\langle \nabla_{a}\Phi
_{2},X^{2}\right\rangle +\left\langle \Phi_{2},\nabla_{a}X^{2}\right\rangle ,
\tag{f}%
\end{equation}
and the result is proved.
\end{proof}

\begin{itemize}
\item $\nabla_{a}$ acting on the duality left contracted product of $\Phi\in%
%TCIMACRO{\dbigwedge }%
%BeginExpansion
{\displaystyle\bigwedge}
%EndExpansion
\mathcal{V}^{\star}(U)$ with $X\in%
%TCIMACRO{\dbigwedge }%
%BeginExpansion
{\displaystyle\bigwedge}
%EndExpansion
\mathcal{V}(U)$ (or, $X\in%
%TCIMACRO{\dbigwedge }%
%BeginExpansion
{\displaystyle\bigwedge}
%EndExpansion
\mathcal{V}(U)$ with $\Phi\in%
%TCIMACRO{\dbigwedge }%
%BeginExpansion
{\displaystyle\bigwedge}
%EndExpansion
\mathcal{V}^{\star}(U)$) satisfies the Leibniz rule, i.e.,
\begin{align}
\nabla_{a}\left\langle \Phi,X\right\vert  &  =\left\langle \nabla_{a}%
\Phi,X\right\vert +\left\langle \Phi,\nabla_{a}X\right\vert ,\label{CDMMF16}\\
\nabla_{a}\left\langle X,\Phi\right\vert  &  =\left\langle \nabla_{a}%
X,\Phi\right\vert +\left\langle X,\nabla_{a}\Phi\right\vert .\label{CDMMF17}%
\end{align}

\item $\nabla_{a}$ acting on the duality right contracted product of $\Phi\in%
%TCIMACRO{\dbigwedge }%
%BeginExpansion
{\displaystyle\bigwedge}
%EndExpansion
\mathcal{V}^{\star}(U)$ with $X\in%
%TCIMACRO{\dbigwedge }%
%BeginExpansion
{\displaystyle\bigwedge}
%EndExpansion
\mathcal{V}(U)$ (or, $X\in%
%TCIMACRO{\dbigwedge }%
%BeginExpansion
{\displaystyle\bigwedge}
%EndExpansion
\mathcal{V}^{^{\star}}(U)$ with $\Phi\in%
%TCIMACRO{\dbigwedge }%
%BeginExpansion
{\displaystyle\bigwedge}
%EndExpansion
\mathcal{V}(U)$) satisfies the Leibniz rule, i.e.,%
\begin{align}
\nabla_{a}\left\vert \Phi,X\right\rangle  &  =\left\vert \nabla_{a}%
\Phi,X\right\rangle +\left\vert \Phi,\nabla_{a}X\right\rangle ,
\label{CDMMF18}\\
\nabla_{a}\left\vert X,\Phi\right\rangle  &  =\left\vert \nabla_{a}%
X,\Phi\right\rangle +\left\vert X,\nabla_{a}\Phi\right\rangle .
\label{CDMMF19}%
\end{align}

\end{itemize}

\begin{proof}
We will prove only the statement given by Eq.(\ref{CDMMF16}). Take $\Phi\in%
%TCIMACRO{\dbigwedge }%
%BeginExpansion
{\displaystyle\bigwedge}
%EndExpansion
\mathcal{V}^{\star}(U),$ $X\in%
%TCIMACRO{\dbigwedge }%
%BeginExpansion
{\displaystyle\bigwedge}
%EndExpansion
\mathcal{V}(U)$ and $\Psi\in%
%TCIMACRO{\dbigwedge }%
%BeginExpansion
{\displaystyle\bigwedge}
%EndExpansion
\mathcal{V}^{\star}(U)$. By a property of the duality left contracted product,
we have%
\begin{equation}
\left\langle \left\langle \Phi,X\right\vert ,\Psi\right\rangle =\left\langle
X,\widetilde{\Phi}\wedge\Psi\right\rangle . \tag{a}%
\end{equation}

Now, by using Eq.(\ref{CDMMF15}) and Eq.(\ref{CDMMF14}), we can write%
\begin{align}
&  \left\langle \nabla_{a}\left\langle \Phi,X\right\vert ,\Psi\right\rangle
+\left\langle \left\langle \Phi,X\right\vert ,\nabla_{a}\Psi\right\rangle
\nonumber\\
&  =\left\langle \nabla_{a}X,\widetilde{\Phi}\wedge\Psi\right\rangle
+\left\langle X,\nabla_{a}(\widetilde{\Phi}\wedge\Psi)\right\rangle
\nonumber\\
&  =\left\langle \nabla_{a}X,\widetilde{\Phi}\wedge\Psi\right\rangle
+\left\langle X,(\nabla_{a}\widetilde{\Phi})\wedge\Psi)\right\rangle
+\left\langle X,\widetilde{\Phi}\wedge\nabla_{a}\Psi)\right\rangle . \tag{b}%
\end{align}
Thus, taking into account Eq.(\ref{CDMMF11}) and by recalling once again a
property of the duality left contracted product, it follows that%
\begin{equation}
\left\langle \nabla_{a}\left\langle \Phi,X\right\vert ,\Psi\right\rangle
=\left\langle \left\langle \Phi,\nabla_{a}X\right\vert ,\Psi\right\rangle
+\left\langle \left\langle \nabla_{a}\Phi,X\right\vert ,\Psi)\right\rangle .
\tag{c}%
\end{equation}

Then, by the non-degeneracy of the duality scalar product, the required result follows.
\end{proof}

\section{Deformed Covariant Derivative}

Let $\left\langle U,\Gamma\right\rangle $ be a parallelism structure on $U,$
and let $\nabla_{a}$ be its associated $a$-DCDO. Take an invertible smooth
extensor\textbf{\ }operator field $\lambda$ on $V\supseteq U.$ Recall that
given $\lambda$, it is possible to construct a deformed parallelism structure
\cite{fmcr2} $\left\langle U,\overset{\lambda}{\Gamma}\right\rangle $ on $U$,
with associated $a$-DCDO denoted by $\overset{\lambda}{\nabla}_{a}.$

Recall \cite{fmcr2} that the deformed covariant derivative operator$\overset
{\lambda}{\text{ }\nabla}_{a}$ has two basic properties, namely, for
all\footnote{Recall that $\lambda^{\bigtriangleup}$ is the so-called
\emph{duality adjoint} of $\lambda,$ and $\lambda^{-\bigtriangleup}$ is a
\emph{short notation} for $(\lambda^{\bigtriangleup})^{-1}=(\lambda
^{-1})^{\bigtriangleup}$.} $v\in\mathcal{V}(U)$,$\overset{\lambda}{\nabla}%
_{a}v=\lambda(\nabla_{a}\lambda^{-1}(v))$ and for $\omega\in\mathcal{V}%
^{\star}(U)$,$\overset{\lambda}{\text{ }\nabla}_{a}\omega=\lambda
^{-\bigtriangleup}(\nabla_{a}\lambda^{\bigtriangleup}(\omega))$.

We give now two properties of $\overset{\lambda}{\text{ }\nabla}_{a}$which are
generalizations of the basic properties just quoted above.

\begin{itemize}
\item For all $X\in%
%TCIMACRO{\dbigwedge }%
%BeginExpansion
{\displaystyle\bigwedge}
%EndExpansion
\mathcal{V}(U)$%
\begin{equation}
\overset{\lambda}{\nabla}_{a}X=\underline{\lambda}(\nabla_{a}\underline
{\lambda}^{-1}(X)), \label{DCD3}%
\end{equation}
where $\underline{\lambda}$ is the so-called \emph{extended} of $\lambda,$ and
$\underline{\lambda}^{-1}$ is a \emph{more simple notation} for $(\underline
{\lambda})^{-1}=\underline{(\lambda^{-1})}.$

\item For all $\Phi\in%
%TCIMACRO{\dbigwedge }%
%BeginExpansion
{\displaystyle\bigwedge}
%EndExpansion
\mathcal{V}^{\star}(U)$%
\begin{equation}
\overset{\lambda}{\nabla}_{a}\Phi=\underline{\lambda}^{-\bigtriangleup}%
(\nabla_{a}\underline{\lambda}^{\bigtriangleup}(\Phi)), \label{DCD4}%
\end{equation}
where $\underline{\lambda}^{\bigtriangleup}=(\underline{\lambda}%
)^{\bigtriangleup}=\underline{(\lambda^{\bigtriangleup})}$ and $\underline
{\lambda}^{-\bigtriangleup}=(\underline{\lambda}^{\bigtriangleup})^{-1}.$
\end{itemize}

\begin{proof}
To prove the property for smooth multivector fields as given by Eq.(\ref{DCD3}%
), we make use of a noticeable criterion:

(\textbf{i}) We check the statement for scalar fields $f\in\mathcal{S}(U)$.
Using Eq.(\ref{CDMMF1}) and recalling a basic property of the extension
procedure \cite{fmcr1}, we have%
\[
\overset{\lambda}{\nabla}_{a}f=af=\underline{\lambda}(af)=\underline{\lambda
}(\nabla_{a}f)=\underline{\lambda}(\nabla_{a}\underline{\lambda}^{-1}(f)).
\]

(\textbf{ii}) Next, we check the statement for simple $k$-vector fields
$v_{1}\wedge\cdots\wedge v_{k}\in%
%TCIMACRO{\tbigwedge \nolimits^{k}}%
%BeginExpansion
{\textstyle\bigwedge\nolimits^{k}}
%EndExpansion
\mathcal{V}(U).$ In this case, we use mathematical induction.

We have that for $k=1,$ the property for vector fields as given by
Eq.(\ref{DCD1}), is true.

For $k>1,$ we must prove the inductive implication%
\begin{align*}
\overset{\lambda}{\nabla}_{a}(v_{1}\wedge\cdots v_{k})  &  =\underline
{\lambda}(\nabla_{a}\underline{\lambda}^{-1}(v_{1}\wedge\cdots v_{k}))\\
\Longrightarrow\overset{\lambda}{\nabla}_{a}(v_{1}\wedge\cdots v_{k}\wedge
v_{k+1})  &  =\underline{\lambda}(\nabla_{a}\underline{\lambda}^{-1}%
(v_{1}\wedge\cdots v_{k}\wedge v_{k+1})).
\end{align*}

The inductive step, follows using Eq.(\ref{CDMMF7}) and recalling a basic
property of the extension procedure. Indeed,
\begin{align*}
\overset{\lambda}{\nabla}_{a}(v_{1}\wedge\cdots v_{k}\wedge v_{k+1})  &
=(\overset{\lambda}{\nabla}_{a}(v_{1}\wedge\cdots v_{k}))\wedge v_{k+1}%
+(v_{1}\wedge\cdots v_{k})\wedge\overset{\lambda}{\nabla}_{a}v_{k+1}\\
&  =\underline{\lambda}(\nabla_{a}\underline{\lambda}^{-1}(v_{1}\wedge\cdots
v_{k}))\wedge\underline{\lambda}\underline{\lambda}^{-1}(v_{k+1})\\
&  +\underline{\lambda}\underline{\lambda}^{-1}(v_{1}\wedge\cdots v_{k}%
)\wedge\underline{\lambda}(\nabla_{a}\underline{\lambda}^{-1}(v_{k+1}))\\
&  =\underline{\lambda}((\nabla_{a}\underline{\lambda}^{-1}(v_{1}\wedge\cdots
v_{k}))\wedge\underline{\lambda}^{-1}(v_{k+1})\\
&  +\underline{\lambda}^{-1}(v_{1}\wedge\cdots v_{k})\wedge\nabla
_{a}\underline{\lambda}^{-1}(v_{k+1}))\\
&  =\underline{\lambda}(\nabla_{a}(\underline{\lambda}^{-1}(v_{1}\wedge\cdots
v_{k})\wedge\underline{\lambda}^{-1}(v_{k+1})))\\
&  =\underline{\lambda}(\nabla_{a}\underline{\lambda}^{-1}(v_{1}\wedge\cdots
v_{k}\wedge v_{k+1})).
\end{align*}

(\textbf{iii}) We now check the statement for\ a finite addition of simple $k
$-vector fields $X^{k}+\cdots Z^{k}\in%
%TCIMACRO{\tbigwedge \nolimits^{k}}%
%BeginExpansion
{\textstyle\bigwedge\nolimits^{k}}
%EndExpansion
\mathcal{V}(U). $ Using Eq.(\ref{CDMMF6}) and recalling the linear operator
character for the extended of a linear operator, we have%
\begin{align*}
\overset{\lambda}{\nabla}_{a}(X^{k}+\cdots Z^{k})  &  =\overset{\lambda
}{\nabla}_{a}X^{k}+\cdots\overset{\lambda}{\nabla}_{a}Z^{k}\\
&  =\underline{\lambda}(\nabla_{a}\underline{\lambda}^{-1}(X^{k}%
))+\cdots\underline{\lambda}(\nabla_{a}\underline{\lambda}^{-1}(Z^{k}))\\
&  =\underline{\lambda}(\nabla_{a}\underline{\lambda}^{-1}(X^{k})+\cdots
\nabla_{a}\underline{\lambda}^{-1}(Z^{k}))\\
&  =\underline{\lambda}(\nabla_{a}(\underline{\lambda}^{-1}(X^{k}%
)+\cdots\underline{\lambda}^{-1}(Z^{k})))\\
&  =\underline{\lambda}(\nabla_{a}\underline{\lambda}^{-1}(X^{k}+\cdots
Z^{k})).
\end{align*}
Hence, it necessarily follows that the statement must be true for all smooth
$k$-vector fields.

(\textbf{iv}) We now can prove the statement for multivector fields $X\in%
%TCIMACRO{\tbigwedge }%
%BeginExpansion
{\textstyle\bigwedge}
%EndExpansion
\mathcal{V}(U)$. Indeed, using Eq.(\ref{CDMMF3}) and Eq.(\ref{CDMMF6}), and
recalling the linear operator character for the extended of a linear operator,
we get%
\begin{align*}
\overset{\lambda}{\nabla}_{a}X  &  =\overset{n}{\underset{k=0}{%
%TCIMACRO{\tsum }%
%BeginExpansion
{\textstyle\sum}
%EndExpansion
}}\overset{\lambda}{\nabla}_{a}X^{k}=\overset{n}{\underset{k=0}{%
%TCIMACRO{\tsum }%
%BeginExpansion
{\textstyle\sum}
%EndExpansion
}}\underline{\lambda}(\nabla_{a}\underline{\lambda}^{-1}(X^{k}))\\
&  =\underline{\lambda}(\nabla_{a}\underline{\lambda}^{-1}(\overset
{n}{\underset{k=0}{%
%TCIMACRO{\tsum }%
%BeginExpansion
{\textstyle\sum}
%EndExpansion
}}X^{k}))=\underline{\lambda}(\nabla_{a}\underline{\lambda}^{-1}(X)),
\end{align*}
and the result is proved.
\end{proof}

\section{Relative Covariant Derivative}

Let $\left\langle U_{0},\Gamma\right\rangle $ be a parallelism structure on
$U_{0},$ and let $\nabla_{a}$ be its associated $a$-\textit{DCDO}. Take any
relative parallelism structure $\left\langle U,B\right\rangle $ compatible
with $\left\langle U_{0},\Gamma\right\rangle $ (i.e., $U_{0}\cap
U\neq\emptyset$)$.$ Denote by $\partial_{a}$ its associated $a$-\textit{DCDO}.
Recall that there exists a well-defined smooth extensor operator field on
$U_{0}\cap U$, which we called the\emph{\ relative connection field}
$\gamma_{a}$, which satisfies the the \emph{split theorem} as formulated for
smooth vector fields and for smooth form fields \cite{fmcr2} i.e.: for
all\footnote{Recall that $\gamma_{a}^{\bigtriangleup}$ is the so-called
\emph{duality adjoint of} $\gamma_{a}$.} $v\in\mathcal{V}(U_{0}\cap U)$,
$\nabla_{a}v=\partial_{a}v+\gamma_{a}(v)$, and for all $\omega\in
\mathcal{V}^{\star}(U_{0}\cap U)$, $\nabla_{a}\omega=\partial_{a}\omega
-\gamma_{a}^{\bigtriangleup}(\omega)$.

We present now the split theorem for smooth multivector fields and for
multiform fields, which are the generalization of the theorem just recalled
above for smooth vector fields and for smooth form fields

\noindent\textbf{Theorem. }(a)\textbf{\ }For all $X\in%
%TCIMACRO{\tbigwedge }%
%BeginExpansion
{\textstyle\bigwedge}
%EndExpansion
\mathcal{V}(U_{0}\cap U)$%
\begin{equation}
\nabla_{a}X=\partial_{a}X+\underset{\smile}{\gamma_{a}}(X), \label{RCD3}%
\end{equation}
where $\underset{\smile}{\gamma_{a}}$ is \emph{generalized}\footnote{See
\cite{fmcr1}, to recall the notion of the generalization procedure.}%
\emph{\ of} $\gamma_{a}$

(b) For all $\Phi\in%
%TCIMACRO{\tbigwedge }%
%BeginExpansion
{\textstyle\bigwedge}
%EndExpansion
\mathcal{V}^{\star}(U_{0}\cap U)$%
\begin{equation}
\nabla_{a}\Phi=\partial_{a}\Phi-\underset{\smile}{\gamma_{a}^{\bigtriangleup}%
}(\Phi), \label{RCD4}%
\end{equation}
where $\underset{\smile}{\gamma_{a}^{\bigtriangleup}}$ is the so-called
generalized of $\gamma_{a}^{\bigtriangleup}$ which as we know coincides with
the so-called duality adjoint of $\underset{\smile}{\gamma_{a}}.$

\begin{proof}
To prove the property for smooth multivector fields as given by Eq.(\ref{RCD3}%
), we use of a noticeable criterion already introduced above.

(\textbf{i}) We check the statement for scalar fields $f\in\mathcal{S}(U).$
Using Eq.(\ref{CDMMF1}) and recalling a basic property of the generalization
procedure, we have%
\[
\nabla_{a}f=af=\partial_{a}f=\partial_{a}f+\gamma_{a}(f).
\]

(\textbf{ii}) Next, we check the statement for simple $k$-vector fields
$v_{1}\wedge\cdots\wedge v_{k}\in%
%TCIMACRO{\tbigwedge \nolimits^{k}}%
%BeginExpansion
{\textstyle\bigwedge\nolimits^{k}}
%EndExpansion
\mathcal{V}(U).$ We use mathematical induction.

For $k=1,$ it is just the property for vector fields which, as we know is true.

For $k>1,$ we have to prove the inductive implication:
\begin{align*}
\nabla_{a}(v_{1}\wedge\cdots v_{k})  &  =\partial_{a}(v_{1}\wedge\cdots
v_{k})+\gamma_{a}(v_{1}\wedge\cdots v_{k})\\
\Longrightarrow\nabla_{a}(v_{1}\wedge\cdots v_{k}\wedge v_{k+1})  &
=\partial_{a}(v_{1}\wedge\cdots v_{k}\wedge v_{k+1})+\gamma_{a}(v_{1}%
\wedge\cdots v_{k}\wedge v_{k+1}).
\end{align*}

The inductive step is assured using Eq.(\ref{CDMMF7}) and recalling a basic
property of the generalization procedure. Indeed,%
\begin{align*}
\nabla_{a}(v_{1}\wedge\cdots v_{k}\wedge v_{k+1})  &  =(\nabla_{a}(v_{1}%
\wedge\cdots v_{k}))\wedge v_{k+1}+(v_{1}\wedge\cdots v_{k})\wedge\nabla
_{a}v_{k+1}\\
&  =(\partial_{a}(v_{1}\wedge\cdots v_{k})+\gamma_{a}(v_{1}\wedge\cdots
v_{k}))\wedge v_{k+1}\\
&  +(v_{1}\wedge\cdots v_{k})\wedge(\partial_{a}v_{k+1}+\gamma_{a}(v_{k+1}))\\
&  =(\partial_{a}(v_{1}\wedge\cdots v_{k}))\wedge v_{k+1}+(v_{1}\wedge\cdots
v_{k})\wedge\partial_{a}v_{k+1}\\
&  +\gamma_{a}(v_{1}\wedge\cdots v_{k})\wedge v_{k+1}+(v_{1}\wedge\cdots
v_{k})\wedge\gamma_{a}(v_{k+1})\\
&  =\partial_{a}(v_{1}\wedge\cdots v_{k}\wedge v_{k+1})+\gamma_{a}(v_{1}%
\wedge\cdots v_{k}\wedge v_{k+1}).
\end{align*}

(\textbf{iii}) Next we check the statement for a finite addition of simple $k
$-vector fields $X^{k}+\cdots Z^{k}\in%
%TCIMACRO{\tbigwedge \nolimits^{k}}%
%BeginExpansion
{\textstyle\bigwedge\nolimits^{k}}
%EndExpansion
\mathcal{V}(U). $ Using Eq.(\ref{CDMMF6}) and recalling the linear operator
character for the generalized of a linear operator, we have%
\begin{align*}
\nabla_{a}(X^{k}+\cdots Z^{k})  &  =\nabla_{a}X^{k}+\cdots\nabla_{a}Z^{k}\\
&  =\partial_{a}X^{k}+\underset{\smile}{\gamma_{a}}(X^{k})+\cdots\partial
_{a}Z^{k}+\underset{\smile}{\gamma_{a}}(Z^{k})\\
&  =\partial_{a}(X^{k}+\cdots Z^{k})+\gamma_{a}(X^{k}+\cdots Z^{k}).
\end{align*}
Hence, it immediately follows that the statement must be true for all smooth
$k$-vector fields.

(\textbf{iv}) We can now prove the statement for multivector fields $X\in%
%TCIMACRO{\tbigwedge }%
%BeginExpansion
{\textstyle\bigwedge}
%EndExpansion
\mathcal{V}(U).$ Using Eq.(\ref{CDMMF3}) and Eq.(\ref{CDMMF6}), and recalling
the linear operator character for the generalized of a linear operator, we get%
\begin{align*}
\nabla_{a}X  &  =\underset{k=0}{\overset{n}{%
%TCIMACRO{\tsum }%
%BeginExpansion
{\textstyle\sum}
%EndExpansion
}}\nabla_{a}X^{k}=\underset{k=0}{\overset{n}{%
%TCIMACRO{\tsum }%
%BeginExpansion
{\textstyle\sum}
%EndExpansion
}(}\partial_{a}X^{k}+\underset{\smile}{\gamma_{a}}(X^{k}))\\
&  =\partial_{a}\underset{k=0}{\overset{n}{%
%TCIMACRO{\tsum }%
%BeginExpansion
{\textstyle\sum}
%EndExpansion
}}X^{k}+\underset{\smile}{\gamma_{a}}(\underset{k=0}{\overset{n}{%
%TCIMACRO{\tsum }%
%BeginExpansion
{\textstyle\sum}
%EndExpansion
}}X^{k})=\partial_{a}X+\underset{\smile}{\gamma_{a}}(X),
\end{align*}
which is what we wanted to prove.\medskip
\end{proof}

Let $\left\langle U,B\right\rangle $ and $\left\langle U^{\prime},B^{\prime
}\right\rangle $, $U\cap U^{\prime}\neq\emptyset$, be two compatible
parallelism structures taken on a smooth manifold $M$. The $a$-\textit{DCDO}'s
associated with $\left\langle U,B\right\rangle $ and $\left\langle U^{\prime
},B^{\prime}\right\rangle $ are denoted by $\partial_{a}$ and $\partial
_{a}^{\prime}$, respectively. As we already know \cite{fmcr2}, there exists a
well-defined smooth extensor operator field on $U\cap U^{\prime}$, the
\emph{Jacobian field} $J,$ such that the following two basic properties are
satisfied: for all $v\in\mathcal{V}(U\cap U^{\prime})$, $\partial_{a}^{\prime
}v=J(\partial_{a}J^{-1}(v))$, and for all $\omega\in\mathcal{V}^{\star}(U\cap
U^{\prime})$, $\partial_{a}^{\prime}\omega=J^{-\bigtriangleup}(\partial
_{a}J^{\bigtriangleup}(\omega))$.

We can see immediately that the basic properties just recalled implies that
$\partial_{a}^{\prime}$ is the $J$-deformation of $\partial_{a}.$

We present now two properties for the relative covariant derivatives which are
generalizations of the basic properties just recalled above.

\begin{itemize}
\item For all $X\in%
%TCIMACRO{\dbigwedge }%
%BeginExpansion
{\displaystyle\bigwedge}
%EndExpansion
\mathcal{V}(U\cap U^{\prime})$%
\begin{equation}
\partial_{a}^{\prime}X=\underline{J}(\partial_{a}\underline{J}^{-1}(X)).
\label{RCD7}%
\end{equation}
It is an immediate consequence of Eq.(\ref{DCD3}).

\item For all $\Phi\in%
%TCIMACRO{\dbigwedge }%
%BeginExpansion
{\displaystyle\bigwedge}
%EndExpansion
\mathcal{V}^{\star}(U\cap U^{\prime})$%
\begin{equation}
\partial_{a}^{\prime}\Phi=\underline{J}^{-\bigtriangleup}(\partial
_{a}\underline{J}^{\bigtriangleup}(\Phi)). \label{RCD8}%
\end{equation}
It is an immediate consequence of Eq.(\ref{DCD4}).
\end{itemize}

\section{Conclusions}

We developed using the algebra of multivectors, multiforms and extensors
\cite{fmcr1} and the theory of parallelism structures on smooth manifolds
\cite{fmcr2} a theory of covariant derivatives, deformed covariant derivatives
an relative covariant derivatives of multivector and multiform fields,
detailing important results.

\end{document}